\documentclass[13pt]{article}
\usepackage{graphicx} 
\usepackage{authblk}
\usepackage{url}
\usepackage{amsfonts} 
\usepackage{amssymb}
\usepackage{mathtools}
\usepackage[utf8]{inputenc}
\usepackage{mathrsfs}
\usepackage{xcolor}

\usepackage{appendix}
\usepackage{geometry}
\usepackage{amsthm}

\theoremstyle{definition}

\theoremstyle{remark}

\title{Charged and rotating boson stars in 5-dimensional\\
Einstein-Maxwell(-Chern-Simons) theory}

\author[1]{Yves Brihaye}
\author[2]{Betti Hartmann}

\affil[1]{Physique de l'Univers-Champs et Gravitation, Universit\'e de
Mons, 7000 Mons, Belgium}
\affil[2]{Department of Mathematics, University College London, Gower Street, London, WC1E 6BT, UK}

\date{17 March 2023}

\begin{document}
\maketitle

\begin{abstract}
We study charged and rotating boson stars in 5-dimensional Einstein-Maxwell(-Chern-Simons) theory
assuming the two angular momenta associated to the two orthogonal planes of rotation to be equal.
Next to the angular momenta, the boson stars carry electric charge and magnetic moment. Interestingly, we find new branches of Einstein-Maxwell-Chern-Simons solutions for which the spatial part of the gauge potential possesses nodes. Consequently, the magnetic moment and the gyromagnetic ratio have opposite sign 
as compared to the solutions on the main branch. For sufficiently large energy density we find that the solutions possess ergoregions.

\end{abstract}

\section{Introduction}
With General Relativity now the accepted and experimentally well confirmed paradigm to describe the gravitational interaction for a wide range of masses and sizes of objects, it remains to be understood how strong gravity acts on scales where quantum effects play an important role. While a consistent theory of Quantum Gravity that would also be able to explain a number of puzzles such as that of dark energy has not be formulated to this day there are possibilities to test 
strong gravity in settings that could be connected to Quantum Theory. One such possibility is the boson star \cite{BS1,BS2,BS3,BS4,kunz_list_kleihaus,kunz_list_kleihaus_schaffer} which is made off a scalar field that is essentially quantum in nature as its collapse is prevented by Heisenberg's uncertainty relation. One could think of such a star as a ``macroscopic Bose-Einstein condensate'' that is self-gravitating. These solutions exist due to a global U(1) symmetry of the model that leads to a conserved Noether charge that can be interpreted as the number of scalar bosonic particles making up the star. These solitonic objects are stationary as they possess a harmonic time-dependence, in their simplest version, however, have a static energy density that leads to a static space-time. Boson stars can also rotate (with resulting stationary space-time)\cite{BS2,BS3,BS4,kunz_list_kleihaus,kunz_list_kleihaus_schaffer} and interestingly the resulting angular momentum is given as an integer multiple of the Noether charge. Hence, the angular momentum is quantized - a feature that is very common in quantum physics - and is proportional to the total number of scalar bosonic particles that make up the star. It has been argued in \cite{ergoregion_BS} that boson stars with large angular momentum possess an ergoregion which would eventually make them unstable. 

Gauging the U(1) symmetry leads to charged boson stars \cite{Jetzer:1989av,Kleihaus:2009kr,Pugliese:2013gsa}. The non-rotating boson stars possess electric charge proportional to the Noether charge with the proportionality constant equal to the gauge coupling. These solutions exist as long as the electromagnetic repulsion does not overcome the gravitational attraction \cite{Jetzer:1989av}, i.e. a critical value of the gauge coupling exists at fixed gravitational coupling.
Adding rotation leads to solutions with electric charge and magnetic moment \cite{Brihaye:2009dx, Collodel:2019ohy}. It was shown in \cite{Collodel:2019ohy} that the relation between angular momentum and Noether charge present in the uncharged case also holds in the presence of a U(1) gauge field.

Boson stars can be constructed as well in higher space-time dimensions which requires a complex scalar field doublet
\cite{Hartmann:2010pm}. In 5 space-time dimensions, rotating stars can possess two angular momenta. Choosing these two angular momenta equal, the symmetry of the system can be enhanced and the space-time possesses hyper-spherical symmetry. 
As for boson stars in 4 space-time dimensions, the sum of the angular momenta is proportional to the Noether charge. 
One aim of this present paper is to add a U(1) gauge field
to the model discussed in \cite{Hartmann:2010pm}. As we will shown below, these solutions possess electric charge and magnetic moment. Next to the standard Maxwell term, another possibility exists in odd space-time dimensions: a Chern-Simons
gauge field interaction. While the former is a relativistic
gauge field model, the Chern-Simons term is topological
and does not depend on the metric. The latter is important when
building models describing phenomena in non-relatistic physics such as e.g. condensed matter. Charged black holes without scalar fields in 
Einstein-Maxwell-Chern-Simons theory have been studied in 
\cite{Kunz:2005ei,kunz:2015kja,Kunz:2017pnm,Blazquez:2017kig}, while 5-dimensional charged, rotating
black holes with scalar hair have been studied in \cite{Brihaye:2018mlv}. Here, we construct the globally regular counterparts to these black holes and extend the results to include a Chern-Simons term.

Our paper is organized as follows: in Section 2 with discuss the model, while Section 3 contains our numerical results. We conclude in Section 4.


\section{The Model}
The action of the model that we will consider in the following reads~:
\begin{equation}
\label{eq:action}
S=\int \left[ \frac{{\cal R}}{16\pi G}
   -  \left( D_\mu \Phi \right)^\dagger \left( D^\mu \Phi \right) - V( \left|\Phi \right|)
   - \frac{1}{4} F_{\mu \nu} F^{\mu \nu} + \alpha \frac{1}{\sqrt{-g}} \epsilon^{\mu \nu \rho \sigma \theta}
	A_{\mu}F_{\nu \rho}F_{\sigma \theta}
 \right] \sqrt{-g} \ {\rm d}^5x \ .
\end{equation}
This is a U(1) gauge field model  coupled minimally to a 
complex scalar doublet $\Phi=(\phi_1,\phi_2)^T$ with potential $V( \left|\Phi \right|)$ as well as Einstein gravity with ${\cal R}$ the Ricci scalar and $G$ Newton's constant.
Note that the scalar sector possesses a global $U(2)$ symmetry whose any $U(1)$ subgroup can be gauged.
Here, the diagonal part of the $U(1)\times U(1)$ maximal Abelian subgroup is gauged. The covariant derivative
and U(1) field strength tensor then take the form
\begin{equation}
    D_\mu = (\partial_\mu - i q A_\mu)  \ \ , \ \ F_{\mu \nu} = \partial_\mu A_{\nu} - \partial_\nu A_{\mu}
\end{equation}
and $q$ denotes the gauge coupling constant. We will assume $q>0$ without loss of generality since the sign of $q$ can be absorbed in the gauge fields and the Chern-Simons coupling $\alpha$. The variation of the action (\ref{eq:action}) with respect to the metric
leads to the Einstein equation:~
\begin{equation}
G_{\mu\nu}= R_{\mu\nu}-\frac{1}{2}g_{\mu\nu}R = 8\pi G (T^s_{\mu\nu} + T^v_{\mu\nu})
\  \label{ee} \end{equation}
with the stress-energy tensor of the scalar field
\begin{eqnarray}
T^{s}_{\mu\nu}  =
   (D_\mu \Phi)^\dagger (D_\nu \Phi) 
  + (D_\nu \Phi)^\dagger (D_\mu \Phi)-  \frac{1}{2} 
   g_{\mu\nu} \bigg[ (D_\alpha \Phi)^\dagger (D_\beta \Phi)
  + (D_\beta \Phi)^\dagger (D_\alpha \Phi)\bigg] g^{\alpha\beta}- 
  g_{\mu\nu}  U(|\Phi|) 
  , \label{tmunu}
\end{eqnarray}
and the stress-energy tensor of the gauge field
\begin{eqnarray}
T^{v}_{\mu\nu} =  - F_{\mu \alpha } F_{\nu}^{\alpha} + \frac{1}{4} g_{\mu\nu} F_{\alpha \beta} F^{\alpha \beta} \ \ ,
\end{eqnarray}
respectively. The variation with respect to the matter fields leads to the equations for the scalar field and gauge field, respectively~:
\begin{eqnarray}
\frac{1}{\sqrt{-g}}
D_\mu\left(\sqrt{-g} D^\mu \Phi \right) =
 \frac{\partial U}{\partial |\Phi|^2} \Phi \ \ ,   \ \ 
\frac{1}{\sqrt{-g}} \partial_{\mu} \sqrt{-g} F^{\mu \nu} = J^{\nu} 
+ 3 \alpha \epsilon^{\nu \rho \sigma \theta \alpha} F_{\rho \sigma} F_{\theta \alpha} 
\label{feqH} 
 \end{eqnarray}
 with the 5-current given by 
 \begin{equation}
 \label{eq:noether_current}
     J^{\nu} = iq ((D^{\nu} \Phi)^\dagger \Phi - \Phi^\dagger (D^{\nu} \Phi)) \ .
 \end{equation}
 Note that the action (\ref{eq:action}) is invariant under a local U(1) transformation up to a divergence, i.e. the equations of motion (\ref{feqH}) are gauge invariant.

\subsection{The Ansatz}
For vanishing gauge field 1-form $A_{\mu} {\rm d}x^{\mu}=0$ the model with action (\ref{eq:action}) was studied first in \cite{Hartmann:2010pm}. We will extend these results here to include electric charge and magnetic moment as well as study the influence of the Chern-Simons term. 

As mentioned above, we assume the solutions to possess bi-azimuthal symmetry,
implying the existence of three commuting Killing vectors,
$\xi = \partial_t$, $\eta_1=\partial_{\varphi_1}$, 
and $\eta_2=\partial_{\varphi_2}$. A suitable metric Ansatz then reads~:
\begin{eqnarray}
\label{metric}
&&
{\rm d}s^2 = -b(r) {\rm d}t^2 + \frac{{\rm d}r^2}{f(r)}
  +   R(r) {\rm d}\theta^2
+h(r)\sin^2\theta \left( {\rm d} \varphi_1 -W(r){\rm d}t \right)^2
+h(r)\cos^2\theta \left( {\rm d} \varphi_2 -W(r){\rm d}t \right)^2 ~~{~~~~~}
\\
\nonumber
&&{~~~~~~}+(  R(r)-h(r))\sin^2\theta \cos^2\theta({\rm d} \varphi_1 -{\rm d} \varphi_2)^2
\end{eqnarray}
where $\theta  \in [0,\pi/2]$, $(\varphi_1,\varphi_2) \in [0,2\pi]$,
and $r$ and $t$ denote the radial and time coordinate, respectively.

For such solutions the isometry group is enhanced from $\mathbb{R} \times U(1)^{2}$
to $\mathbb{R} \times U(2)$. This is nothing else but stating that the two angular momenta $J_1$, $J_2$ associated to rotations by $\varphi_i$, $i=1,2$ are equal to each other $J_1=J_2\equiv J/2$, where $J$ is the total angular momentum. 
The symmetry enhancement mentioned above in particular allows to factorize the angular dependence
and thus leads to ordinary differential equations. The Ansatz for the scalar field then reads \cite{Hartmann:2010pm}~:
\begin{equation}
 \Phi = \phi(r) e^{i \omega  t} 
 \left( \begin{array}{c} 
   \sin\theta  e^{i \varphi_1} \\ \cos\theta  e^{i \varphi_2} 
 \end{array} \right) 
 , \label{phi} 
\end{equation} 
where the frequency $\omega$ parametrises the harmonic time-dependence. 
For the scalar field potential we restrict our study to the simplest case of a massive, non self-interacting scalar field, i.e. we set
   \begin{eqnarray}
   V(|\Phi|) =\mu^2 \Phi^\dagger \Phi  =\mu^2 \phi(r)^2
   \end{eqnarray}
where $\mu$ corresponds to the scalar field mass.

Finally, the Ansatz for the electromagnetic potential is chosen to be~:
\begin{equation}
A_{\mu} {\rm d}x^{\mu} = V(r) {\rm d}t + A(r) (\sin^2(\theta) {\rm d} \varphi_1 + \cos^2(\theta) {\rm d} \varphi_2 )
\end{equation} 
which turns out to be consistent with the symmetries of the metric and scalar fields. The non-vanishing components of the field
strength tensor are then
\begin{equation}
       F_{rt}=\frac{{\rm d}V(r)}{{\rm d}r} \ \ , \ \ F_{r \varphi_1} = \frac{{\rm d}A(r)}{{\rm d}r}  \sin^2 \theta \ \ , \ \  F_{r \varphi_2} = \frac{{\rm d}A(r)}{{\rm d}r}  \cos^2 \theta \ \ , \ \ F_{\theta \varphi_1} = - F_{\theta \varphi_2} = A(r) \sin(2\theta) \ ,
\end{equation}
i.e. our solutions possess electric and magnetic fields. 

Without fixing a metric gauge, a straightforward computation
leads to the following reduced action for the system~:
 \begin{eqnarray}
\label{Leff}
{\cal S}_{\rm eff}=\int {\rm d}r {\rm d}t ~L_{\rm eff},~~~{\rm with~~~~}
L_{\rm eff}=L_g+16 \pi G  (L_{s}+ L_{v} + \alpha L_{cs}),
\end{eqnarray}
\begin{eqnarray}
L_{g}&=&
\sqrt{\frac{fh}{b}}
\bigg(
b'R'+\frac{R}{2h}b'h'+\frac{b}{2R}R'^2+\frac{b}{h}R'h'+\frac{1}{2}Rh W'^2+\frac{2b}{f}\left(4-\frac{h}{R}\right)
\bigg),
\\
 L_{s}&=& R\sqrt{\frac{bh}{f}}
\left[
f\phi'^2+\left(\frac{2}{R}+\frac{(1-q A)^2}{h}-\frac{(\omega-W+q(V+WA))^2}{b}+\mu^2\right)\phi^2
\right] \ , \\
L_{v} &=& R\sqrt{\frac{bh}{f}}  \left( \frac{2 A^2}{R^2} + \frac{f}{2 h} (A')^2 - \frac{f}{2b}(V'+W A')^2 \right)  \ ,
\\
L_{CS} &=& 16 A (A' V - A V') \ ,
\end{eqnarray}
the effective gravity ($g$), scalar field ($s$), gauge field ($v$) and Chern-Simons (CS) Lagrangian density, respectively. 
The prime now and in the following denotes the derivative with respect to $r$.
The equations of motion can then be consistently obtained from this reduced action by varying with respect to 
$h$, $b$, $f$, $R$, $W$, $F$, $V$ and $A$. Note that the effective CS Lagrangian density does not depend on the metric functions and hence will not source the space-time curvature. 

The metric gauge freedom can be fixed afterwards, leading to a system of seven independent
equations plus a constraint which is
a consequence of the other equations.
For the construction of the solutions, we have fixed the metric gauge by taking  
\begin{eqnarray}
R(r)=r^2 
\end{eqnarray}
consistently with the standard analytic form of the Myers-Perry solution \cite{mp}.
Appropriate combinations of the equations can be
used such that
the equation  for $f(r)$ is first order while the equations of the six other functions are second order. 
We hence need a total of thirteen conditions at $r=0$ and/or at $r=\infty$
to specify a boundary value problem.

\subsection{Asymptotic behaviour and boundary conditions}
Boson stars are globally regular solutions. At $r=0$ we impose the following boundary conditions~:
\begin{equation}
      f(0) = 1 \ \ ,  \ b'(0) = 0 \ \ ,  \  h(0) = 0 \ \ ,   \ W'(0) = 0 \ \ ,  \ V(0)=0 \ \ ,  \  A(0)=0 \ , \  A'(0) = 0 \ \  ,  \ \phi(0) = 0 \ .
\label{regular_cond}
\end{equation}
Note that the condition $V(0)=0$ does not result from the requirement of regularity, but is a choice. This can be made without loosing generality
since the equations depend only on the combination $qV+\omega$.

Moreover, we want the solutions to be asymptotically flat, i.e. we require~:
\begin{eqnarray}
\nonumber
&&b(r)=
1
+\frac{{\cal M}}{r^2}+ \dots,
~~f(r)=
1
+\frac{{\cal M}}{r^2}+\dots,
~~h(r)=r^2
+\frac{{\cal V}}{r^2}+\dots,
~~W(r)=\frac{{\cal J}}{r^4}+\dots
\\
\label{eq:inf1}
&&
V(r) = V_{\infty} + \frac{q_e}{r^2}+ \dots , A(r) =  \frac{q_m}{r^2}+ \dots
~~
\phi(r)= c_0\frac{e^{-r\sqrt{\mu^2-(\omega-q V_{\infty})^2}}}{r^{3/2}}+\dots,~~ 
\end{eqnarray}
where ${\cal M}$, ${\cal V}$, ${\cal J}$, $q_e$, $q_m$, $V_{\infty}$ and $c_0$ are free parameters that can only be computed from the numerical solution. 

Note that the asymptotic behaviour of the scalar field tells us that it acquires an effective mass $m_{\rm eff}$ with
\begin{equation}
\label{eq:mass}
m_{\rm eff}^2 \equiv  \mu^2 - (\omega - q V_{\infty})^2 \  =  \ (\mu - \omega + q V_{\infty})(\mu + \omega - q V_{\infty}) \ \ .
\end{equation}
The parameter $V_{\infty}$, i.e. the value of the electric potential $V(r)$ at $r\rightarrow \infty$ turns out to be negative in our numerical calculations. Since $V(0)=0$, the value of $V_{\infty}$ corresponds to the potential difference between the origin and infinity. With the choice $q \geq 0$, this tells us that the first factor on the right-hand side of (\ref{eq:mass}), which we define as 
\begin{equation}
    \Omega:=\mu-\omega +q V_{\infty}
\end{equation}
determines whether the boson star is an exponentially localized solution. Obviously, we need to require $\Omega \geq 0$. For 
$(\omega - q V_{\infty})^2 \geq \mu^2$ we are above the threshold of producing scalar particles of mass $\mu$.

\subsection{Physical quantities}
Before we discuss the relevant physical quantities of the solutions and how they can be extracted from the numerical results we obtain, let us remark that although there are {\it a priori} three (four) parameters 
to be varied in the Maxwell (respectively Chern-Simons) case, Newton's constant $G$ and the mass $\mu$ of the scalar field can be set to unity without loss of generality. This is achieved by a suitable rescaling of the matter fields and the coordinate $r$. This means that we are left with the gauge coupling $q$ in the Maxwell case
and additionally with $\alpha$ in the Maxwell-Chern-Simons case. \\

The mass $M$ and total angular momentum $J=J_1+J_2$ of the solutions have been discussed in \cite{Hartmann:2010pm}. Hence, we just state the expressions here without explicitly deriving them. They read~: 
\begin{eqnarray}
\label{global-charges}
M=-\frac{3\pi }{8 G}{\cal M},~~
J=\frac{\pi}{4 G}{\cal J},
\end{eqnarray}
where ${\cal M}$ and ${\cal J}$ are given in (\ref{eq:inf1}).

Since the model we are discussing here possesses a global U(1) symmetry, there exists an associated locally conserved Noether current. This is the current given in (\ref{eq:noether_current}). The globally conserved Noether charge then is~:
\begin{equation}
         Q = \int \sqrt{- g} \ J^0 {\rm d}^4 x = q N \ \ {\rm with} \ \ N = 2\pi^2 \int_0^{\infty} r^3 \sqrt{\frac{h}{fb}} (\omega + W - q (V+A W)) \phi^2  {\rm d} r\ .
\end{equation}
$N$ can then be interpreted as the total number of bosonic particles making up the boson star and $Q$ as the total charge of $N$ individual particles that each carry charge $q$. Also note that there is a relation between the angular momentum $J$ and $N$ given by \cite{Hartmann:2010pm}
\begin{equation}
\vert J\vert = N \ . 
\end{equation}

We can also define the electric charge $Q_e$ and the magnetic moment $Q_m$, respectively, as follows
\begin{equation} 
Q_e = \frac{\pi}{G} q_e \ \ \ , \ \ \ Q_m = \frac{\pi}{G} q_m \ \ , 
\end{equation} 
where $q_e$ and $q_m$ are given in (\ref{eq:inf1}). Using the equation (\ref{feqH}) it can be shown that 
$Q=Q_e$ as expected. In the numerical calculation the validity of this equality is a good cross-check.
Finally, the gyromagnetic ratio $\gamma$ of our solutions reads
\begin{equation}
\gamma=\frac{2 M Q_m}{Q_e J} = \frac{2 M q_m}{q_e J}  \ .
\end{equation}
\\
We will also need the Ricci scalar ${\cal R}$ in the following. This reads
\begin{eqnarray}
  {\cal R}(r) &=& -f\left(\frac{b''}{b} +\frac{h''}{h} +\frac{2 R''}{R}\right) + \frac{f}{2}\left( \frac{(b')^2}{b^2} + \frac{(h')^2}{h^2} + \frac{(R')^2}{R^2} +\frac{h}{b}(W')^2\right) \nonumber \\
    &+& - \frac{R'}{Rbh}(f b h)' - \frac{1}{2 b h} (f' b h' + f' b' h + f b' h') + \frac{2}{R^2} (4 R - h) \ .
\label{eq:ricci}
\end{eqnarray}
We will see in the discussion of the numerical results that some configurations reach limiting solutions with $b(0) \to  0$ (while $b''(0)$ is finite) suggesting that the scalar curvature at the origin diverges for these solutions. \\

\section{Numerical results}
Due to the non-linearity of the field equations, we have solved the equations numerically using the collocation solver COLSYS \cite{COLSYS}. The obtained solutions typically have accuracy of $10^{-6}$. With appropriate rescalings of the fields and coordinates, we can set $8\pi G\equiv 1$, $\mu\equiv 1$, i.e.
the only two parameters to vary in the following are $q$ and $\alpha$.

\subsection{Einstein-Maxwell (EM) boson stars}
Let us first discuss the solutions in the absence of the Chern-Simons interaction, i.e. for $\alpha = 0$.\\

As a crosscheck of our numerics and to emphasize the changes that the presence of the gauge field brings to the model, let us briefly discuss the case $q=0$ that has been studied in detail in \cite{Hartmann:2010pm}.
We have constructed uncharged, rotating boson star solutions and studied their properties by varying the parameter $\phi'(0)$.  For $\phi'(0)=0$ the scalar field is trivial $\phi(r)\equiv 0$, $\omega=1$ and the space-time is simply 5-dimensional
Minkowski space-time. Note, however, that 
the limit $\phi'(0) \to 0$ is subtle~: while the scalar function $\phi(r)$ becomes trivial, the mass $M$ and Noether charge $N$ do not approach zero in this limit.
In fact, a mass gap forms. This has been discussed in \cite{Hartmann:2010pm}.

\begin{figure}[ht!]
\begin{center}
{\includegraphics[width=8cm]{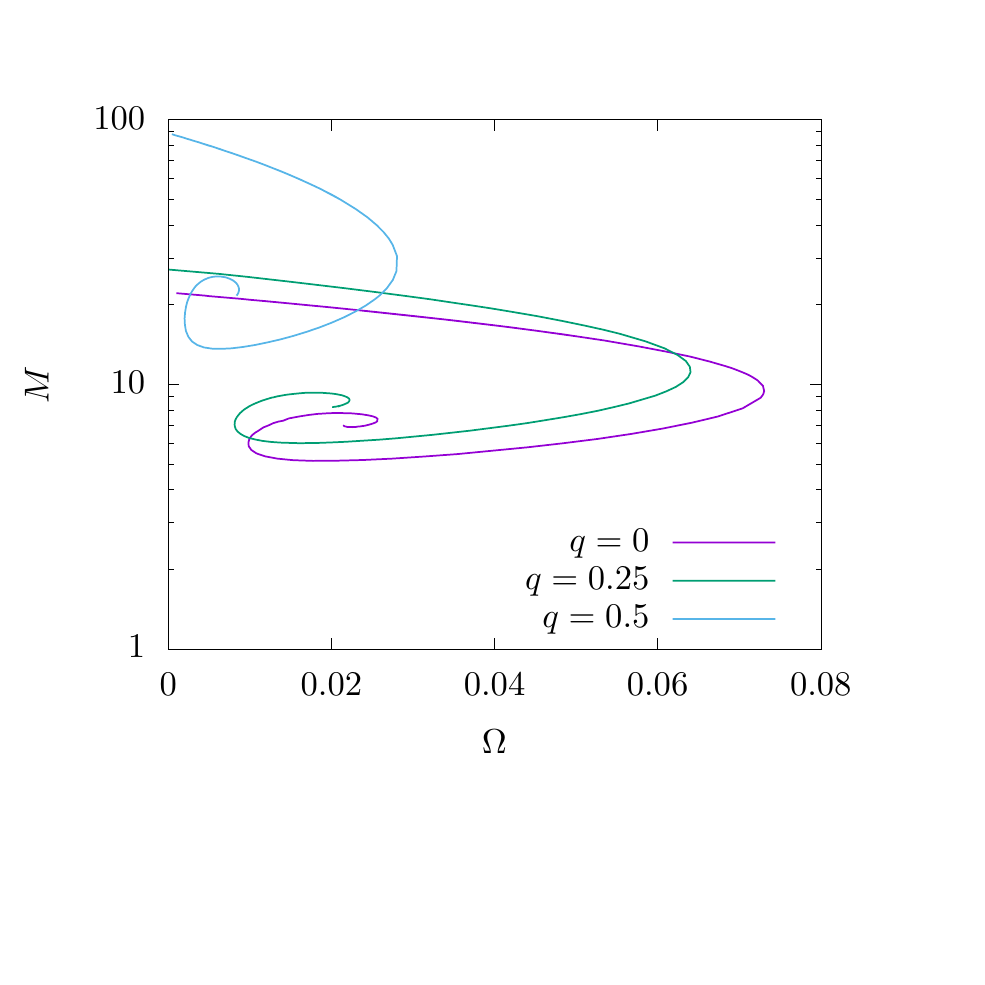}}
{\includegraphics[width=8cm]{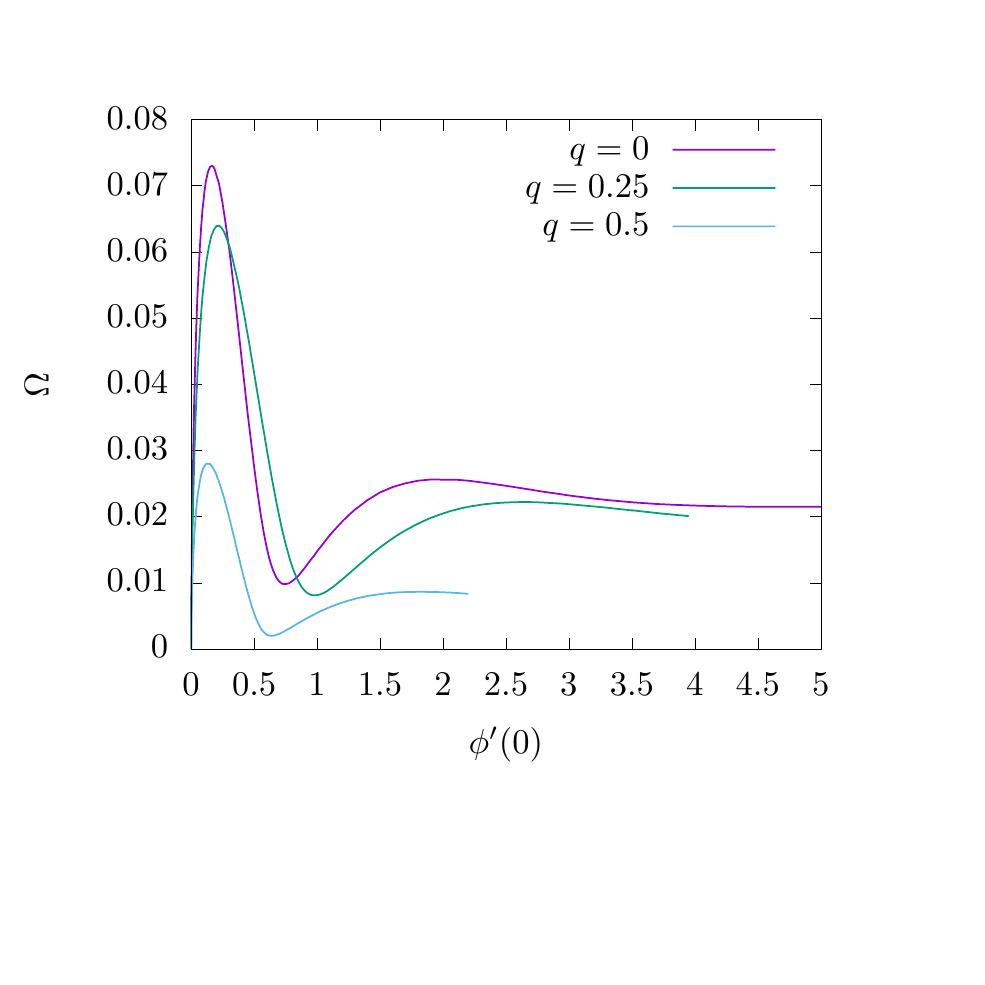}}
\vspace{-2cm}
\end{center}
\caption{The mass $M$ in dependence of $\Omega$ (left) and $\Omega$ in dependence of $\phi'(0)$ (right) for 
EM boson stars with $q=0.25$ (purple) and $q=0.5$ (green), respectively. For comparison we also show the uncharged case, $q=0$.
\label{fig:gamma_Omega_alpha_0}
}
\end{figure}

In Fig. \ref{fig:gamma_Omega_alpha_0} (left) we show the dependence of the mass $M$ on $\Omega$  for $q=0$, $q=0.25$ and $q=0.5$, respectively. 
For all values of $q$ we observe the typical spiraling behaviour, i.e. the existence of a main branch of solutions which exists between $\Omega=0$ and a maximal
value of $\Omega=\Omega_{\rm max,1}$. From $\Omega_{\rm max,1}$ a second branch of solutions extends backwards in $\Omega$ down to $\Omega_{\rm min,2} > 0$. From $\Omega_{\rm min,2}$ a third branch exists up to $\Omega_{\rm max,3} < \Omega_{\rm max,1}$ and bends backwards into a fourth branch.
We find that $\Omega=\Omega_{\rm max,1}$, $\Omega_{\rm min,2}$, $\Omega_{\rm max,3}$ all decrease with increasing $q$, i.e. the interval
in $\Omega$ for which charged, rotating boson stars exist in $5$ dimensions decreases with increasing $q$. Moreover, we find that the mass gap described above for uncharged solutions also exists for charged solutions and increases with increasing $q$.
In fact, we observe that the mass range for which charged boson stars exist changes only slightly when increasing $q$ from zero to $q=0.25$, while the increase to $q=0.5$ increases the value of the mass gap considerably. This is related to the increased
electromagnetic repulsion. The charge $Q$ (and with it the angular momentum $J$) have a very similar qualitative dependence, this is why we do not show them here.

\begin{figure}[ht!]
\begin{center}
{\includegraphics[width=8cm]{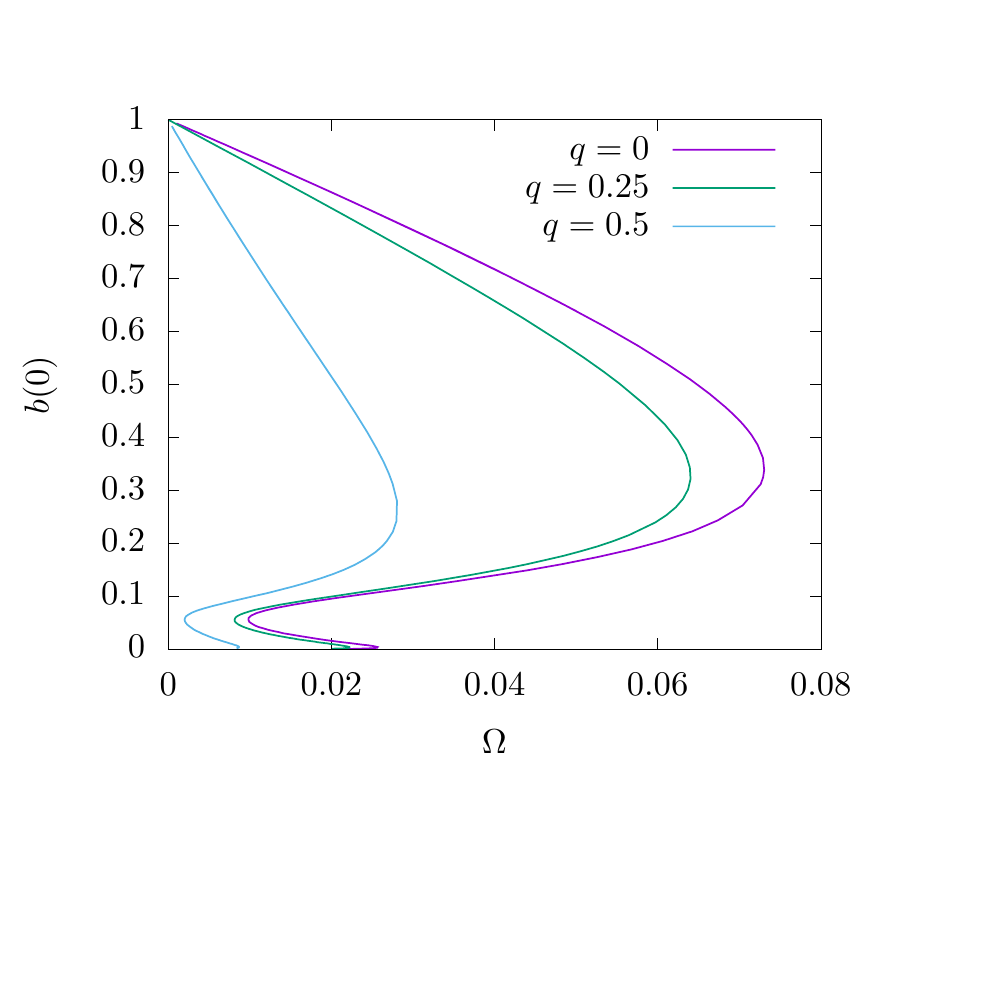}}
{\includegraphics[width=8cm]{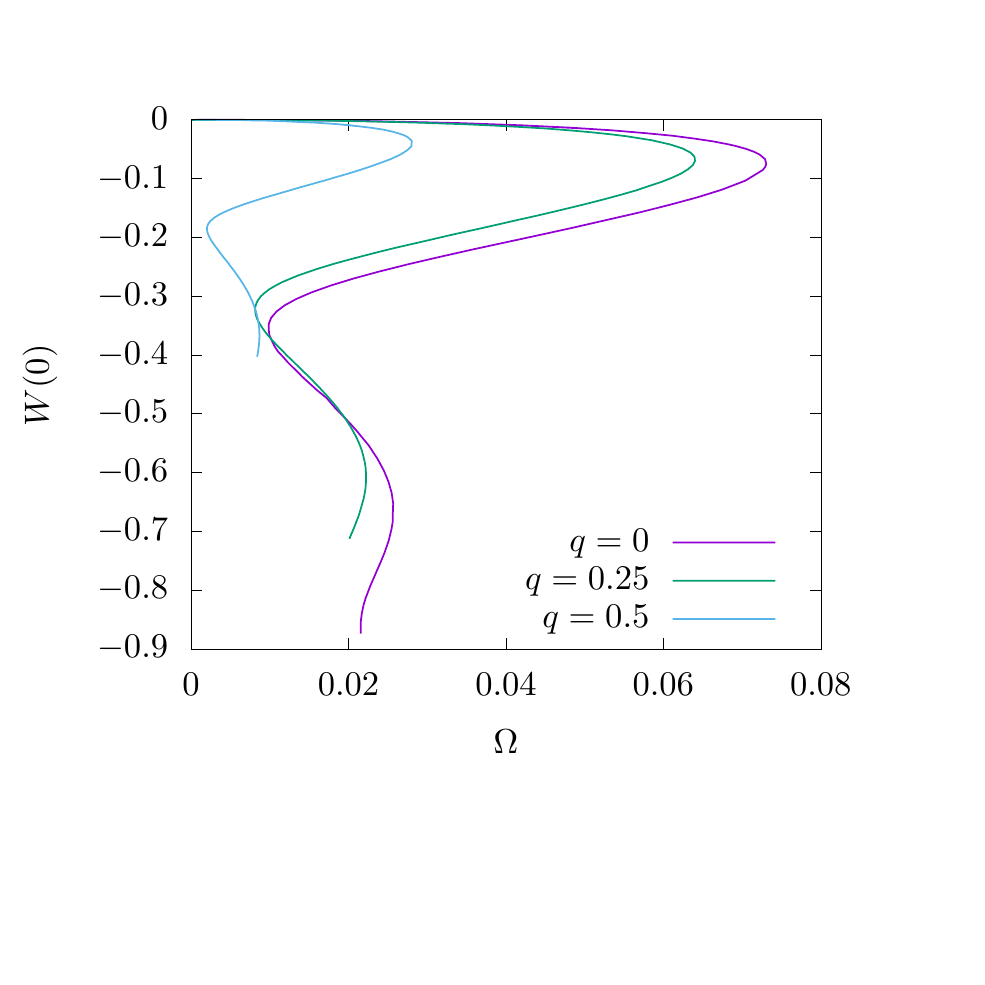}}
\vspace{-2cm}
\end{center}
\caption{The value of the metric function $b(r)$ at the origin, $b(0)$  in dependence of $\Omega$ (left) and the value of the metric function $W(r)$ at the origin, $W(0)$  in dependence of $\Omega$ (left) for 
EM boson stars with $q=0.25$ (purple) and $q=0.5$ (green), respectively. For comparison we also show the uncharged case, $q=0$.
\label{fig:b0_W0_Omega_alpha_0}
}
\end{figure}

Along the branches, the parameter $\phi'(0)$ is increased. We show the dependence of $\Omega$ on $\phi'(0)$ for $q=0$, $q=0.25$ and $q=0.5$ in 
Fig. \ref{fig:gamma_Omega_alpha_0} (right). For $\phi'(0)=0$ we have $\Omega=0$ independent of the choice of $q$. Increasing $\phi'(0)$ the value of $\Omega$ reaches a maximal value, then decreases to a minimal value and for sufficiently large $\phi'(0)$ tends to a constant value of $\Omega$. The solutions
cease to exist when $\phi'(0)$ is too large, where the maximal possible value of $\phi'(0)$ decreases with increasing $q$.
This is related to the formation of a singularity in the Ricci scalar at the origin. The Ricci scalar at $r=0$ is given by ${\cal R}(0) = -4 b''(0)/b(0) -6 f''(0)$ (compare (\ref{eq:ricci})). In Fig. \ref{fig:b0_W0_Omega_alpha_0} (left) we show the value of $b(0)$ in dependence of $\Omega$ for $q=0$, $q=0.25$ and $q=0.5$. For $\Omega=0$ we find
$b(0)=1$ independent of $q$ as this is the limit of vanishing scalar field $\phi(r)\equiv 0$. Along the branches, i.e. increasing $\phi'(0)$, the value of $b(0)$ decreases until it reaches zero, i.e. a solution with diverging Ricci scalar at $r=0$ is reached. We also observe that $W(0)$ decreases from zero when moving along the branches, see
Fig. \ref{fig:b0_W0_Omega_alpha_0} (right). In particular, we find that $g_{tt} = - b + h W^2$ can become zero and even positive indicating that an ergoregion
exists for solutions with sufficiently large $\phi'(0)$.  As has been discussed in the context of boson stars before \cite{ergoregion_BS}, this would make the solutions unstable. We find that the larger $q$, the smaller is the value of $\phi'(0)$ at which an ergoregion appears, e.g. for
$q=0.25$, we find solutions with ergoregions for $\phi'(0) > 1.2$ , while these ergoregions exist for  $\phi'(0) > 0.9$ when choosing $q=0.5$.
Some data is shown in Table \ref{table:ergo_EM}, where we give the two values of $r$ at which $g_{tt}$ becomes zero. The ergoregion is a hyper-spherical
shell of inner radius $r_1$ and outer radius $r_2$. Within this shell, $g_{tt}$ is positive and attains its maximal value at $r^{\rm (max)}$. Values of the maximal
value of $g_{tt}$ and the value of $r^{\rm (max)}$ are also given in Table \ref{table:ergo_EM}. 

\begin{figure}[ht!]
\begin{center}
{\includegraphics[width=8cm]{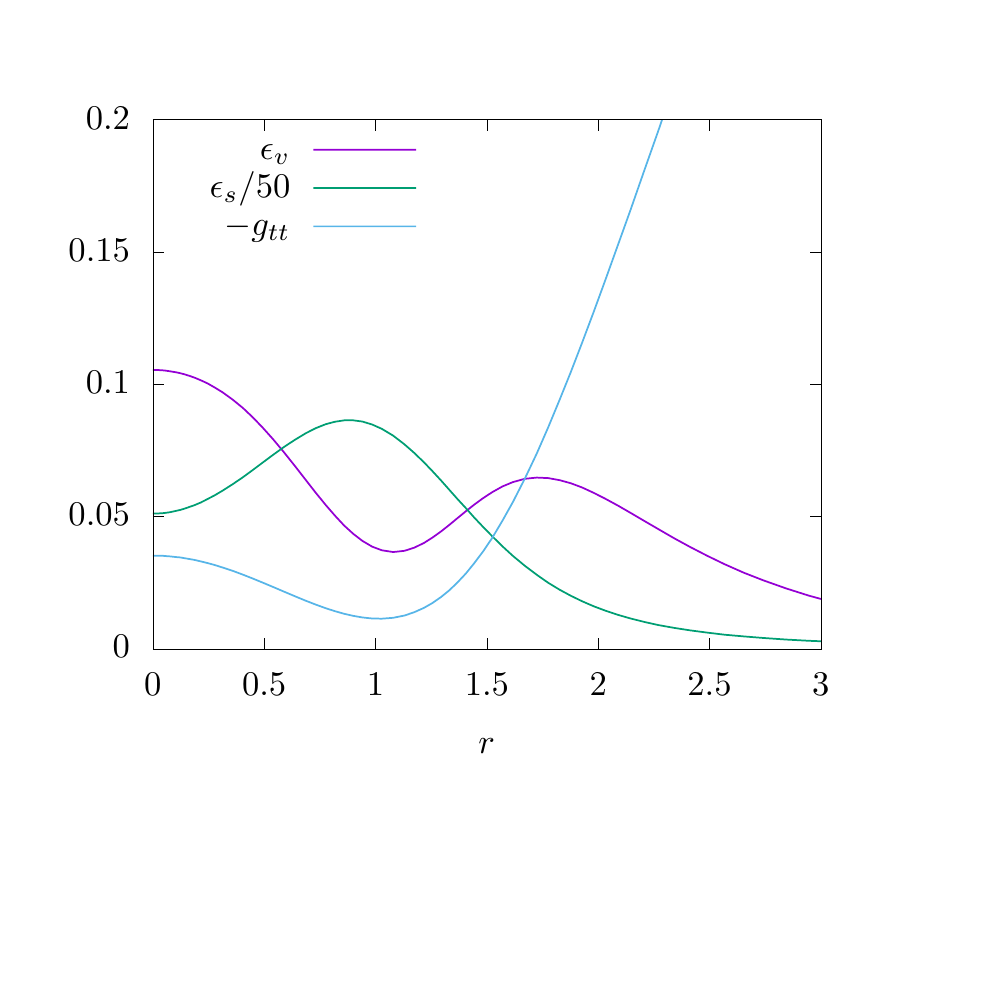}}
{\includegraphics[width=8cm]{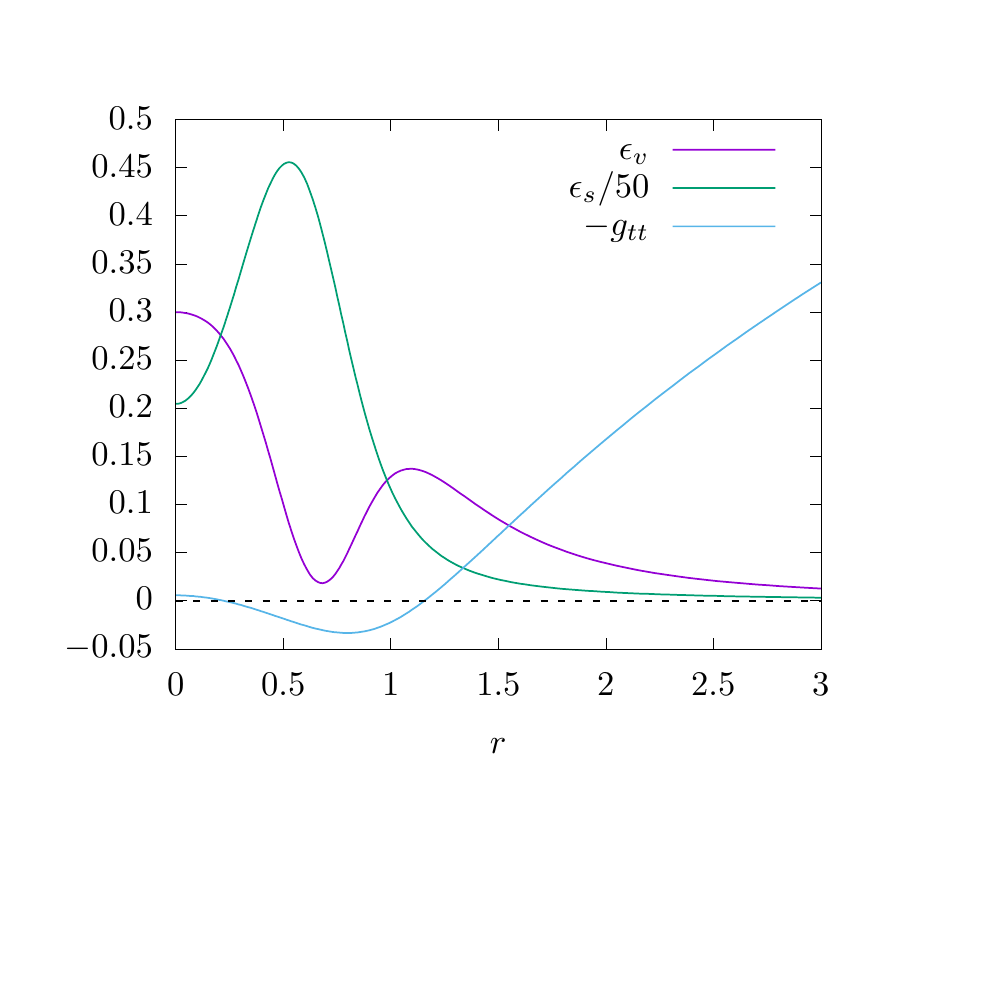}}
\vspace{-2cm}
\end{center}
\caption{The profiles of the gauge field energy density $\epsilon_v$, the scalar field energy density $\epsilon_s$ and the metric function $-g_{tt}$ for EM boson stars with
$q=0.5$, $\gamma=0$ and $\phi'(0)=0.8$ (left) and $\phi'(0)=1.6$ (right). 
\label{fig:energy_density_gamma0_q_0_5}
}
\end{figure}

In Fig. \ref{fig:energy_density_gamma0_q_0_5} we show the gauge field energy density $\epsilon_v$ and scalar field energy density $\epsilon_s$ given by
\begin{eqnarray}
\ \ \epsilon_v \equiv (T^0_0)_v = \frac{f}{2bh}(A')^2(b-hW^2) + \frac{2}{r^4}A^2 + \frac{f}{2b} (V')^2 
\end{eqnarray}
and
\begin{eqnarray}
 \epsilon_s \equiv (T^0_0)_s =  f (\phi')^2 + \mu^2 \phi^2 + \phi^2\left(\frac{2}{r^2} + \frac{(1-qA)^2}{h}\right) + \frac{\phi^2}{b} \left((qV-\omega)^2- W^2(qA-1)^2\right)  \  ,
\end{eqnarray}
respectively. The sum $\epsilon_v +\epsilon_s$ is equivalent to the total energy density of the solution. These profiles are for $q=0.5$ and $\phi'(0)=0.8$ (left) and $\phi'(0)=1.6$ (right), respectively. We also show the metric tensor
component $-g_{tt}$. We observe that the scalar field energy density $\epsilon_s$ dominates the energy density as it is a factor of $50$ larger than the contribution $\epsilon_v$ from the gauge field. The gauge field energy density $\epsilon_v$ is maximal at the center of the boson star, while $\epsilon_s$ has its maximal value at $r=r_{s,{\rm max}} > 0$. Interestingly, the gauge field energy density as well as $-g_{tt}$ have a local minimum around $r_{s,{\rm max}}$. For sufficiently large scalar field energy density 
we find that $-g_{tt}$ becomes negative, i.e. an ergoregion appears. Increasing $\phi'(0)$ from $0.8$ to $1.6$ leads to an increase
of $\Omega$, i.e. the scalar field falls off quicker. Correspondingly, $r_{s,{\rm max}}$ decreases and the maximum of the energy density increases with increasing $\phi'(0)$.

\begin{figure}[ht!]
\begin{center}
{\includegraphics[width=8cm]{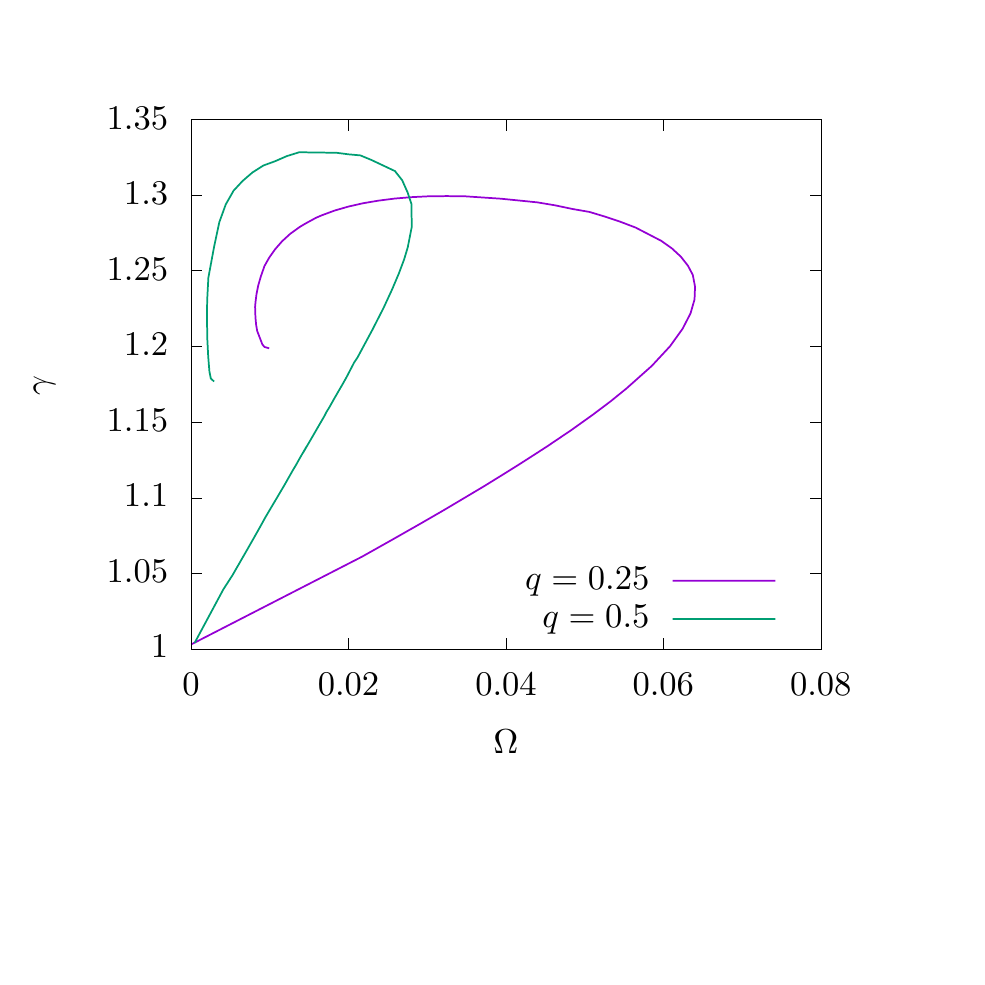}}
{\includegraphics[width=8cm]{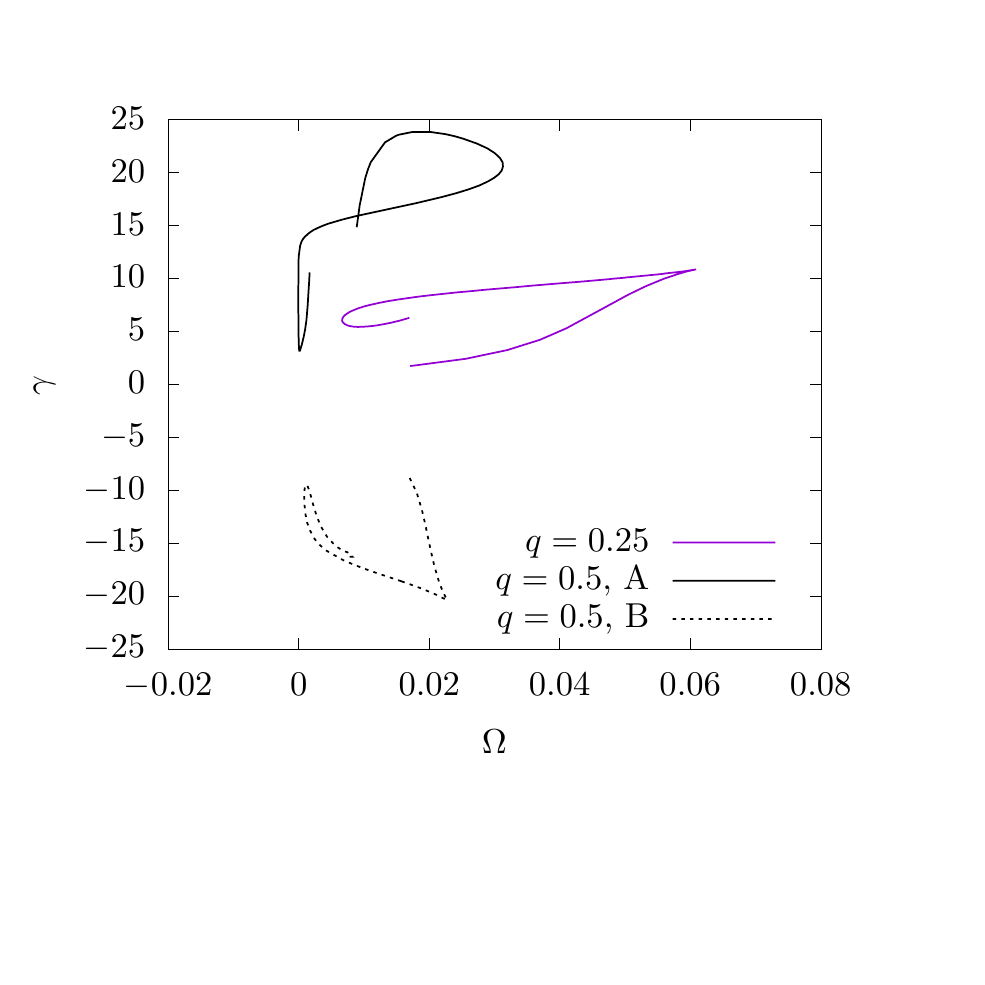}}
\vspace{-2cm}
\end{center}
\caption{{\it Left:} The gyromagnetic ratio $\gamma$  in dependence of $\Omega$ for 
EM boson stars with $q=0.25$ (purple) and $q=0.5$ (green), respectively. {\it Right:} The gyromagnetic ratio $\gamma$  in dependence 
of $\Omega$ for 
EMCS boson stars with $\alpha=1$ and $q=0.25$ (purple) and $q=0.5$ (black), respectively.  In the latter case, we show
branch A (solid) and branch B (dashed), respectively. 
\label{fig:gamma_Omega_alpha_0_1}
}
\end{figure}

We have also studied the gyromagnetic ratio for the solutions. Our results for $q=0.25$ and $q=0.5$ are shown in Fig. \ref{fig:gamma_Omega_alpha_0_1} (left).
On the main branch for  $\Omega\rightarrow 0$ the gyromagnetic ratio tends to the ''classical'' value $\gamma=1$. Increasing $\Omega$ from zero increases $\gamma$ up to a maximal value on the second branch of solutions. The larger $q$ the larger is this maximal value.

\begin{table}[h!]
\begin{center}
\begin{tabular}{ |c|c|c|c|c|c| } 
\hline
$\phi^{\prime}(0)$ & $q$ & $r_1$ & $r_2$ & $r^{{\rm (max)}}$ & $g_{tt}^{{\rm (max)}}$ \\
\hline \hline
 1.6  &  0.25  &    0.29   &    0.97  &   0.75  &  0.038 \\
   1.2   & 0.25  &    0.64    &    0.91  &  0.84  &  0.007 \\
   \hline
1.6  &   0.5 &  0.22   &    1.14   &   0.79     &    0.033\\
  1.0 &  0.5  &  0.66    &  1.18     & 0.90     &    0.008 \\
  0.95 & 0.5   & 0.76    &  1.14   &   0.96      &  0.004 \\
\hline
\end{tabular}
\end{center}
\caption{The two values of $r$ for which $g_{tt}$ becomes zero, i.e. the inner radius $r_1$ and outer radius $r_2$, respectively, of the ergoregion as well as the value of $r=r^{{\rm (max)}}$ at which $g_{tt}$ attains its maximal value $g_{tt}^{{\rm (max)}}$ are given for EM boson stars and some values of $\phi^{\prime}(0)$ and $q$. \label{table:ergo_EM}}
\end{table}

\begin{figure}[ht!]
\begin{center}
{\includegraphics[width=8cm]{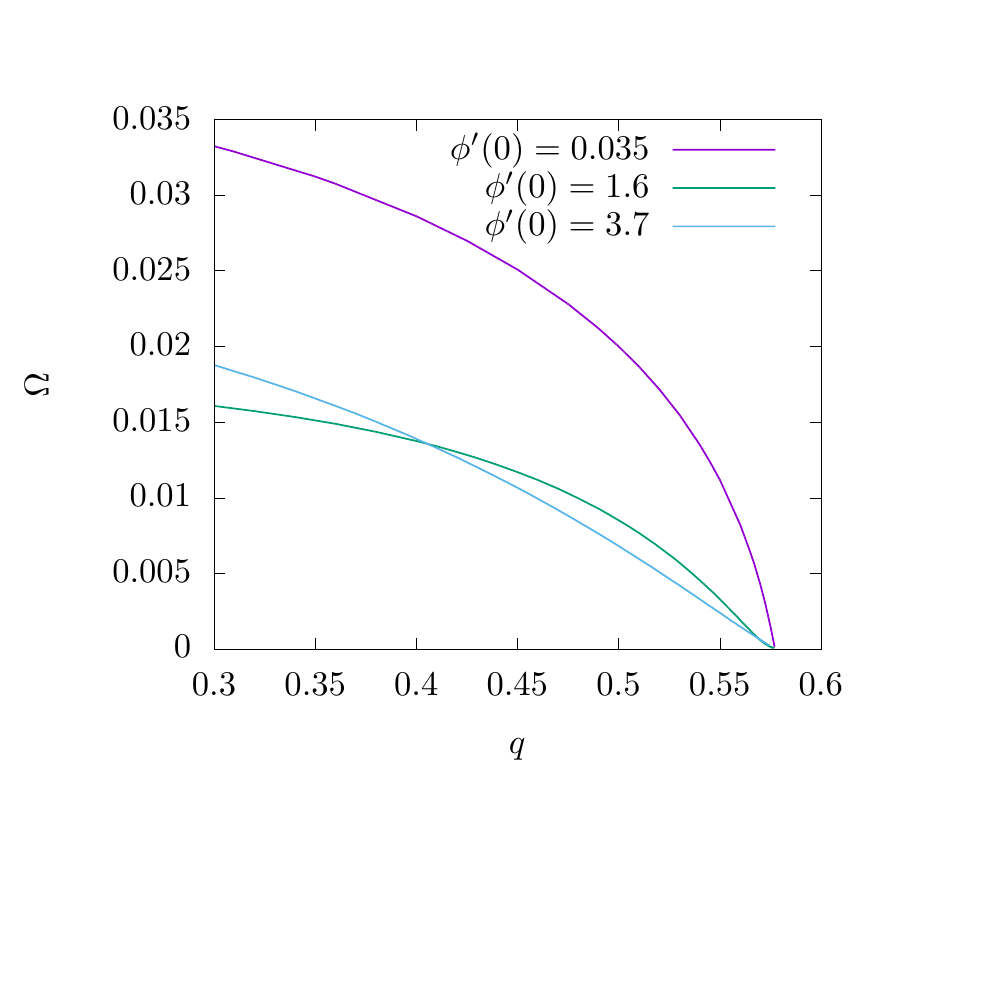}}
{\includegraphics[width=8cm]{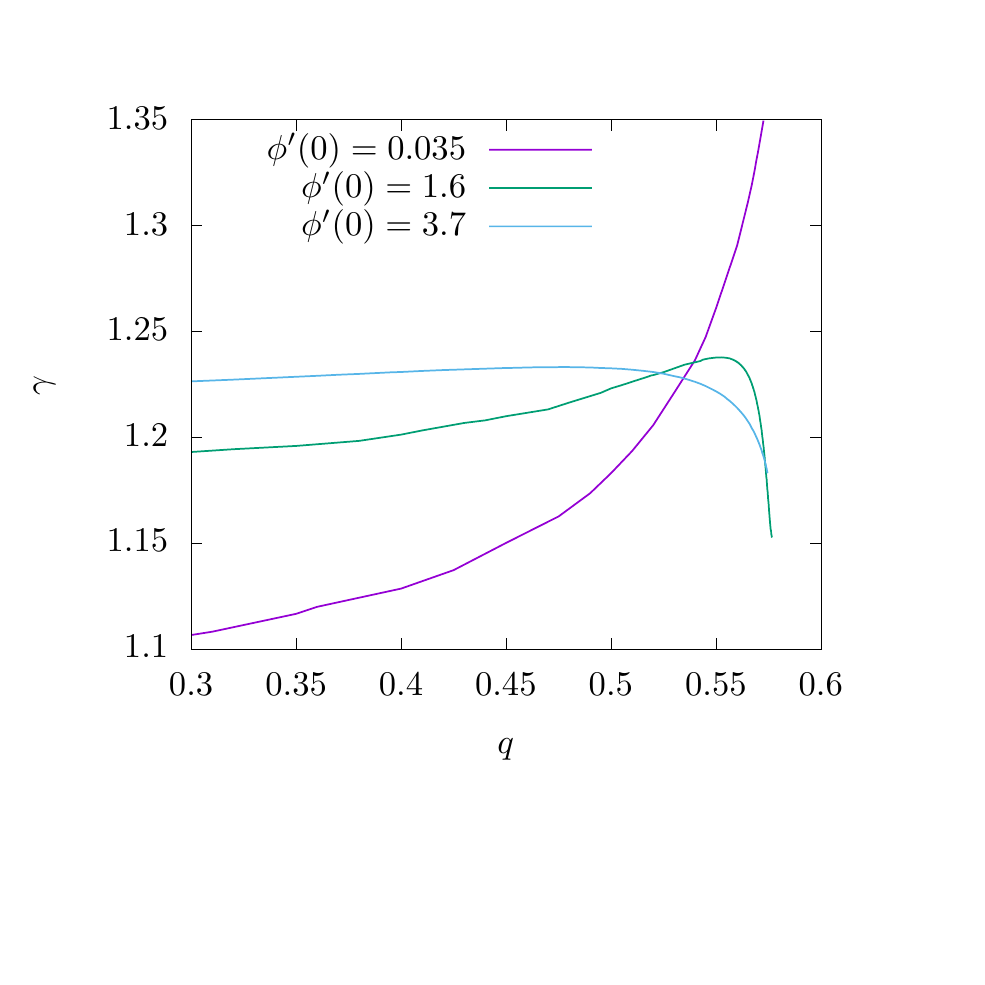}}
\vspace{-2cm}
\end{center}
\caption{{\it Left:} The value of $\Omega$  in dependence of $q$ for 
EM boson stars with $\phi^{\prime}(0)=0.035$ (purple), $\phi^{\prime}(0)=1.6$ (green) and $\phi^{\prime}(0)=3.7$ (blue), respectively. {\it Right:} The gyromagnetic ratio $\gamma$  in dependence 
of $q$ for the same solutions.
\label{fig:gamma_Omega_vs_q_phip_vary}
}
\end{figure}

In order to better understand the dependence of the solutions on $\phi'(0)$ and $q$, we have also studied the case of fixed $\phi'(0)$ and varying $q$.
Our numerical experiments indicate that localized solutions do not exist for $q > q_{\rm max}$ with $q_{\rm max}\approx 0.5775$ more or less independent
of the choice of $\phi'(0) > 0$.  This is shown in Fig. \ref{fig:gamma_Omega_vs_q_phip_vary} (left) where we give $\Omega$ in function of $q$ for
three different values of $\phi'(0)$. Obviously, when $q\rightarrow q_{\rm max}$ the value of $\Omega \rightarrow 0$, i.e. the boson star solution
is no longer (exponentially) localized. Accordingly, all physical quantities subject to a Gauss law (mass $M$, electric charge $Q_e$, magnetic moment $Q_m$, angular momentum $J$) will diverge in this limit. However, it is interesting to note that the gyromagnetic ratio $\gamma$ behaves differently in the limit $q_{\rm max}\approx 0.5775$ when choosing $\phi'(0)$ small as compared to  choosing $\phi'(0)$ large. This is shown in Fig. \ref{fig:gamma_Omega_vs_q_phip_vary} (right) where we give $\gamma$ in function of $q$ for different values of  $\phi'(0)$. We observe that for small values of  $\phi'(0)$ (here  $\phi'(0)=0.035$)
the gyromagnetic ratio increases strongly for $q\rightarrow q_{\rm max}$, while for large values of $\phi'(0)$  (here  $\phi'(0)=1.6$ and  $\phi'(0)=3.7$, respectively) $\gamma$ decreases strongly in this limit. Note that this limit for $q$ was also observed for the corresponding black hole solutions \cite{Brihaye:2018mlv}.

\subsection{Einstein-Maxwell-Chern-Simons (EMCS) boson stars}

\begin{table}[h!]
\begin{center}
\begin{tabular}{ |c|c|c|c|c| } 
\hline
$\phi^{\prime}(0)$ & $\alpha$ & $r_1$ & $r_2$ & branch  \\
\hline \hline
1.6 & 0.5 & 0.33 & 0.87 & A \\
1.6 & 0.5 & 0.31 & 0.93 & B \\
\hline
1.3  &    1.0 & 0.59   &    0.80   &  A \\
1.2  &   1.0 &  0.60    &    0.87  &  B \\
\hline
\end{tabular}
\end{center}
\caption{The two values of $r$ for which $g_{tt}$ becomes zero, i.e. the inner radius $r_1$ and outer radius $r_2$, respectively, of the ergoregion  for EMCS boson stars with $q=0.5$ and some exemplary values of $\phi'(0)$ and $\alpha$. \label{table:ergo_ECSM}}
\end{table}

\begin{figure}[h!]
\begin{center}
{\includegraphics[width=8cm]{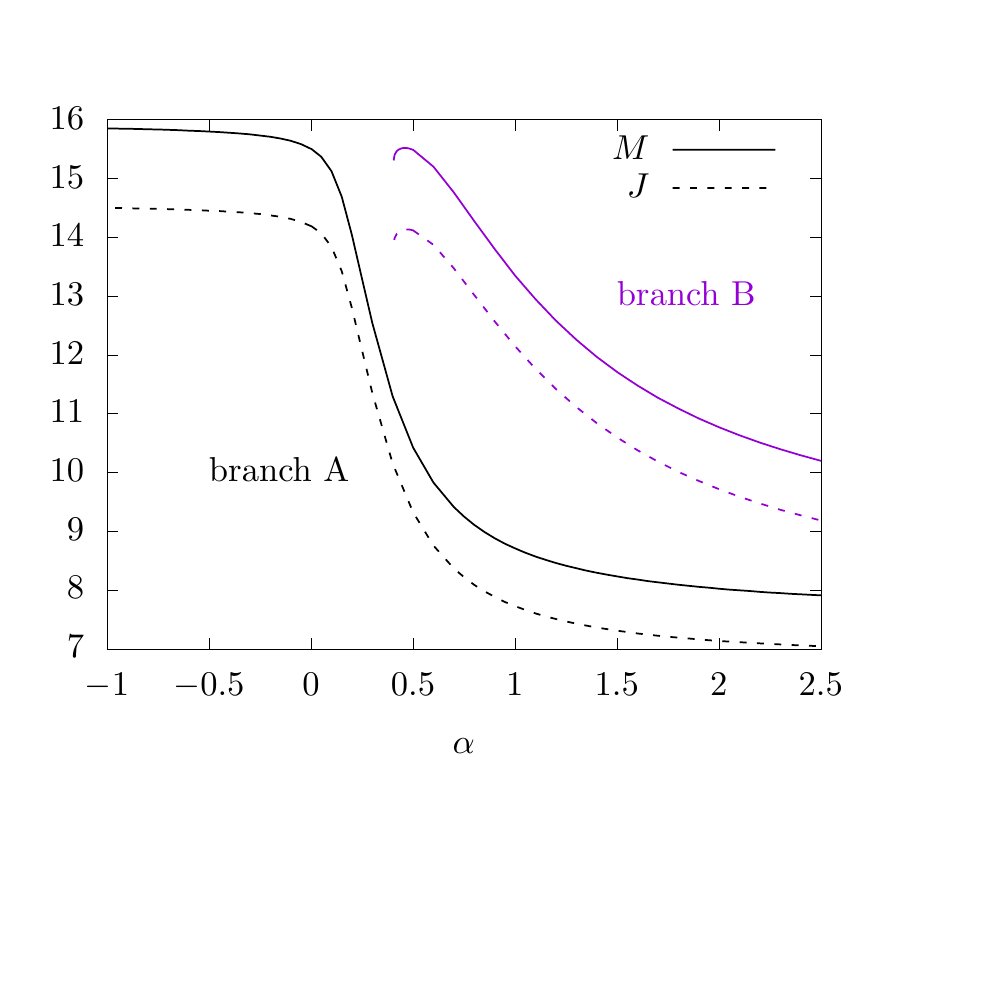}}
{\includegraphics[width=8cm]{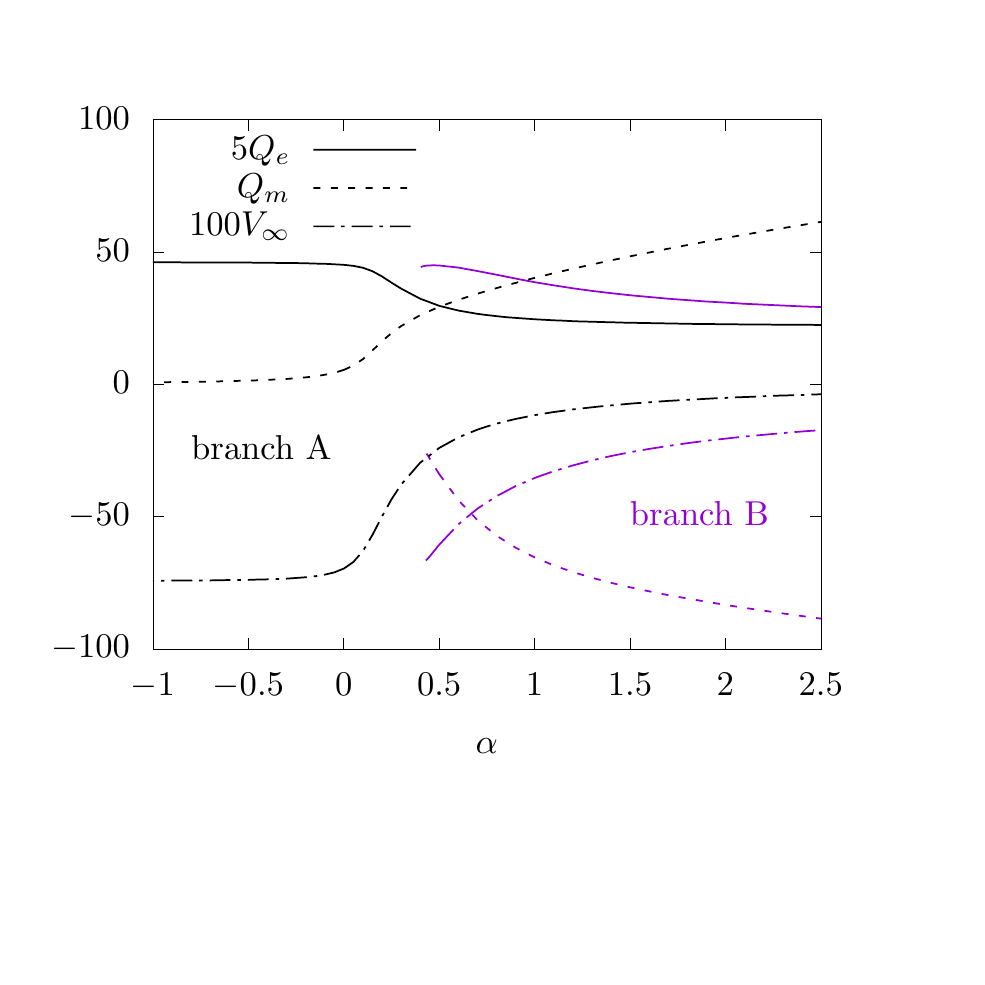}}
\end{center}
\vspace{-2cm}
\caption{The mass $M$ and angular momentum $J$ (left) and the electric charge $Q_e=Q$, the magnetic moment $Q_m$ and the value of the electric potential at infinity $V_{\infty}$ (right) of EMCS boson stars in dependence of $\alpha$ for $\phi'(0)=0.35$ and $q=0.5$. We show branch A with no node (black) and branch B with one node (violet) of the gauge field function $A(r)$, see also Fig. \ref{fig:CS_branchA_branchB}. 
\label{fig:branchA_branchB_CS_compare}
}
\end{figure}

In the following, we discuss the influence of the CS term on the properties of the charged boson stars. 
As expected, the EM boson stars get progressively deformed when choosing $\alpha \neq 0$.  As an example
we show the dependence of the mass $M$ and the angular momentum $J$ (left) as well as the electric charge $Q_e=Q$, the magnetic moment $Q_m$, 
and the value of the electric potential at infinity $V_{\infty}$ (right) on $\alpha$ for $\phi'(0)=0.35$ and $q=0.5$ in Fig. \ref{fig:branchA_branchB_CS_compare}. As these figures suggest, two branches of solutions exist which we will refer to as
'branch A' and 'branch B', respectively, in the following. Branch A is connected to the EM limit $\alpha\rightarrow 0$ and exists for both positive and negative values of $\alpha$, while branch B appears only for sufficiently large and positive values of $\alpha$, i.e.
for $\alpha > \alpha_{\rm cr,B}$, where $\alpha_{\rm cr,B}$ depends on $q$ and $\phi'(0)$.  For $q=0.5$ and $\phi'(0)=0.35$ we find that
$\alpha_{\rm cr,B}\approx 0.405$.

Solutions on both branches have the feature that $M > J$ and both $M$ and $J$ decrease with increasing $\alpha$ 
(except close to $\alpha_{\rm cr,B}$ on branch B where our numerical results indicate an increase on a small interval of $\alpha$). 

\begin{figure}[h!]
\begin{center}
{\includegraphics[width=8cm]{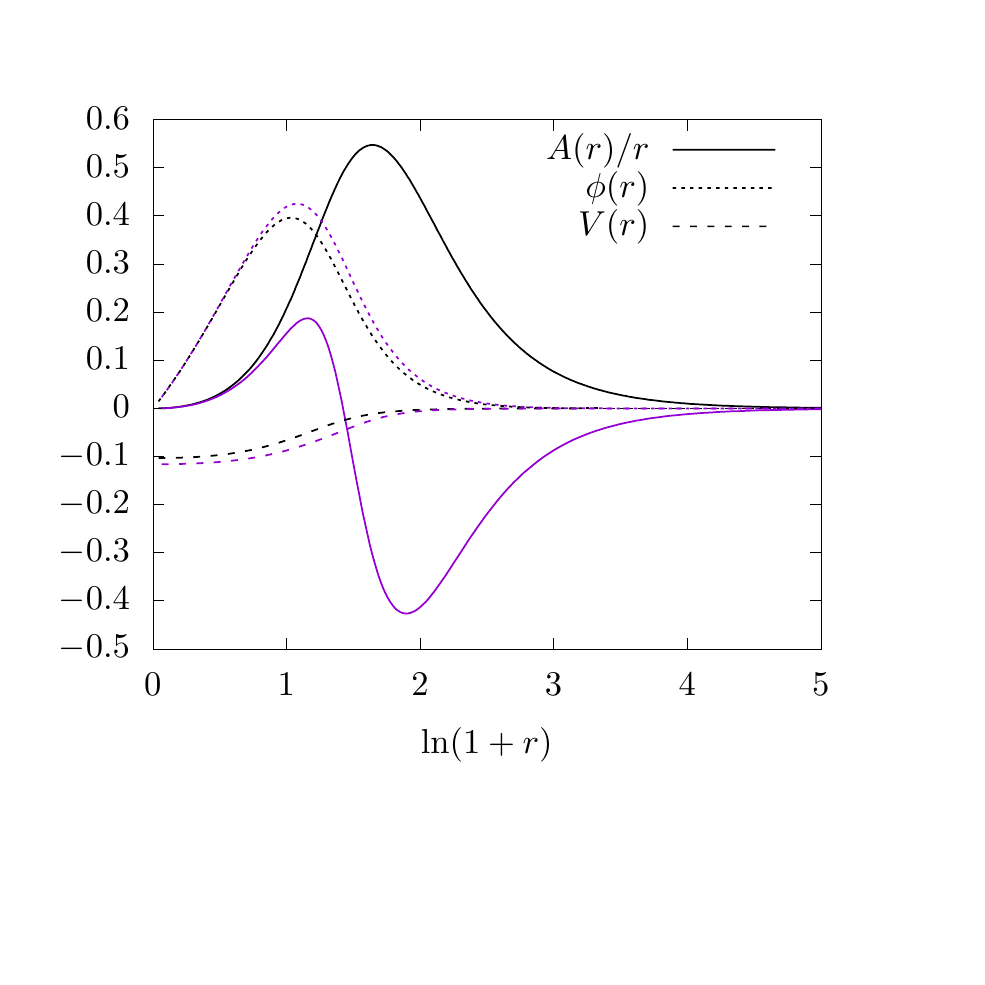}}
{\includegraphics[width=8cm]{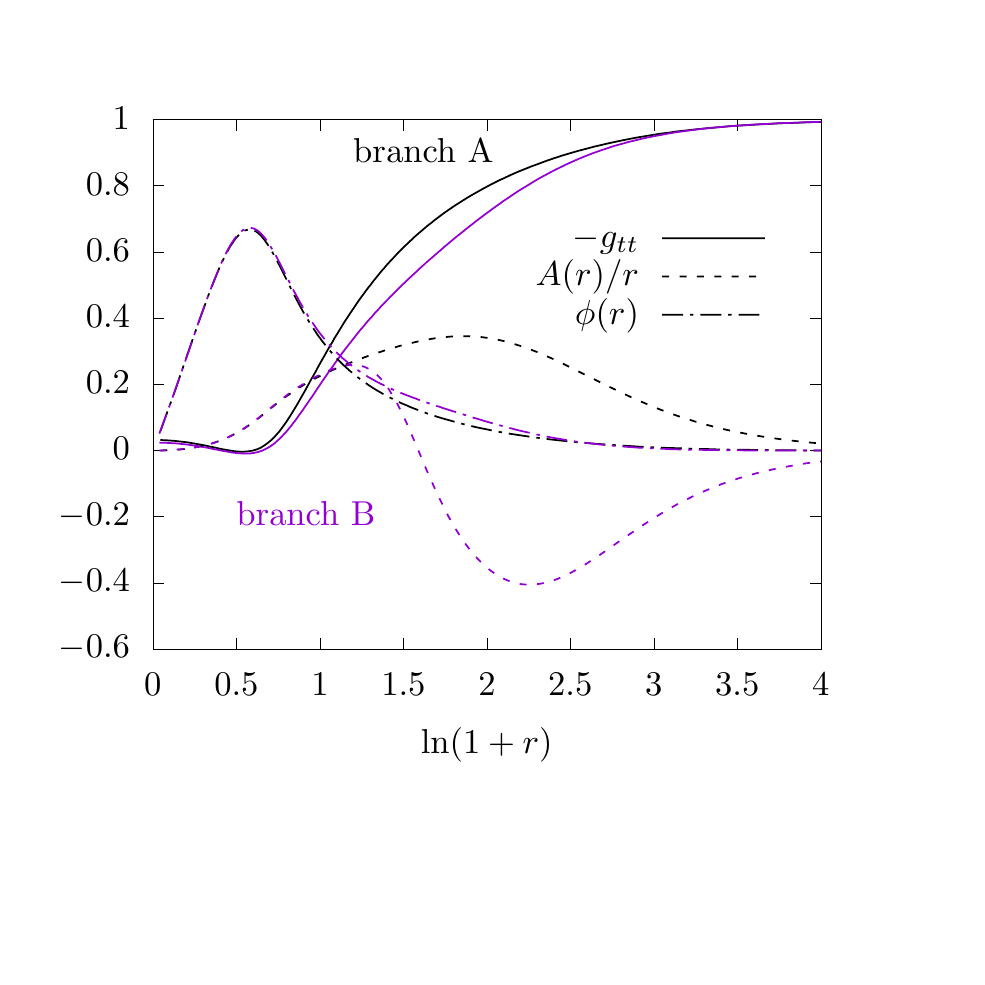}}
\end{center}
\vspace{-2cm}
\caption{{\it Left:} We show the profiles of the gauge potential functions $A(r)/r$ (solid) and $V(r)$ (dashed) as well as of the scalar field function $\phi(r)$ (dotted) for EMCS boson stars on branch A (black) and branch B (violet)
for $q=0.5$, $\alpha = 0.5$ and $\phi'(0)=0.35$. {\it Right:} We show the the metric tensor component $-g_{tt}$ (solid), the gauge potential function $A(r)/r$ (dashed)
and the scalar field function $\phi(r)$ (dotted-dashed) for EMCS boson stars on branch A (black) and branch B (violet)
for $q=0.5$, $\alpha = 0.5$ and $\phi'(0)=1.3$.
\label{fig:CS_branchA_branchB}
}
\end{figure}

The results suggest that mass and angular momentum of the solutions on branch A change little when increasing  $\alpha$ from negative values to zero.  Even for small positive values of $\alpha$, this seems to be the case.
For $\alpha$ slightly smaller, but close to $\alpha_{\rm cr,B}$ we find that $M$ and $J$ drop sharply and then again on a large interval of (positive) $\alpha$ remain nearly constant. This suggests that the appearance of branch B seems to be connected to 
a drop in energy and angular momentum of the boson stars on branch A. All our numerical results indicate that these
two branches remain separated and do not merge at sufficiently large $\alpha$. 
We also observe that the electric charge $Q_e=Q$ decreases with increasing $\alpha$ for both branches (see Fig. \ref{fig:branchA_branchB_CS_compare} (right)) with $Q_e=Q$ smaller on branch A as compared to on branch B. $\vert V_{\infty}\vert $  decreases with increasing $\alpha$ and is - again - larger on branch B. Finally, the magnetic moment $Q_m$ is close to zero 
for negative $\alpha$ (branch A) and increases to positive values when increasing $\alpha$ from zero. On branch B, the magnetic moment is negative and decreases in absolute value when increasing $\alpha$ and approaches zero for large positive values of $\alpha$. The solutions on branch B hence have larger electric charge and larger absolute value of the magnetic moment with the latter being negative on branch B.  In order to understand the difference between the two branches, we have plotted the profiles of typical boson star solutions. This is shown in Fig. \ref{fig:CS_branchA_branchB} (left) for  $q=0.5$, $\alpha = 0.5$ and $\phi'(0)=0.35$.
Clearly, the magnetic potential $A(r)$ possesses a node for the solutions on branch B. Solutions with nodes in the spatial part of the gauge field have been found before for black holes in EMCS theory (without scalar fields) \cite{kunz:2015kja}. These have been interpreted as radial excitations and the fact that solutions on branch B have larger mass than those on branch A
suggests that this interpretation is also suitable here. Interestingly, we observe that neither the electric part of the gauge potential 
(given in terms of the function $V(r)$) nor the scalar field function $\phi(r)$ are strongly changed when radially
exciting the magnetic part of the gauge potential. Fixing $q$ and $\alpha$ and increasing $\phi'(0)$ we find that the 
value of $r=r_0$ at which $A(r)$ becomes zero increases. This is shown in Fig. \ref{fig:CS_branchA_branchB} (right) for $\phi'(0)=1.3$.
In this case we find that for $0 \leq r \lesssim r_0$ the solutions on the two branches barely differ from each other. 
This includes the extend and existence of the ergoregion which is slightly more extended for solutions on branch B, see also Table \ref{table:ergo_ECSM} for some more data. This data also suggests that at fixed $\phi'(0)$ and fixed $q$ the ergoregion has smaller radial thickness when the CS term is present. 

The solutions on branch B possess negative magnetic moment and hence negative gyromagnetic ratio. Our results for $\alpha=1$ and $q=0.25$ as well as $\alpha=1$, $q=0.5$ and branch A and branch B are shown in Fig.\ref{fig:gamma_Omega_alpha_0_1} (right). In comparison to the EM case, the gyromagnetic ratio can become negative (branch B for $q=0.5$) and significantly larger in absolute value. The maximal possible value of $\gamma$ increases with $q$ and is one order of magnitude larger as compared to the EM boson stars.

\begin{figure}[h!]
\begin{center}
{\includegraphics[width=8cm]{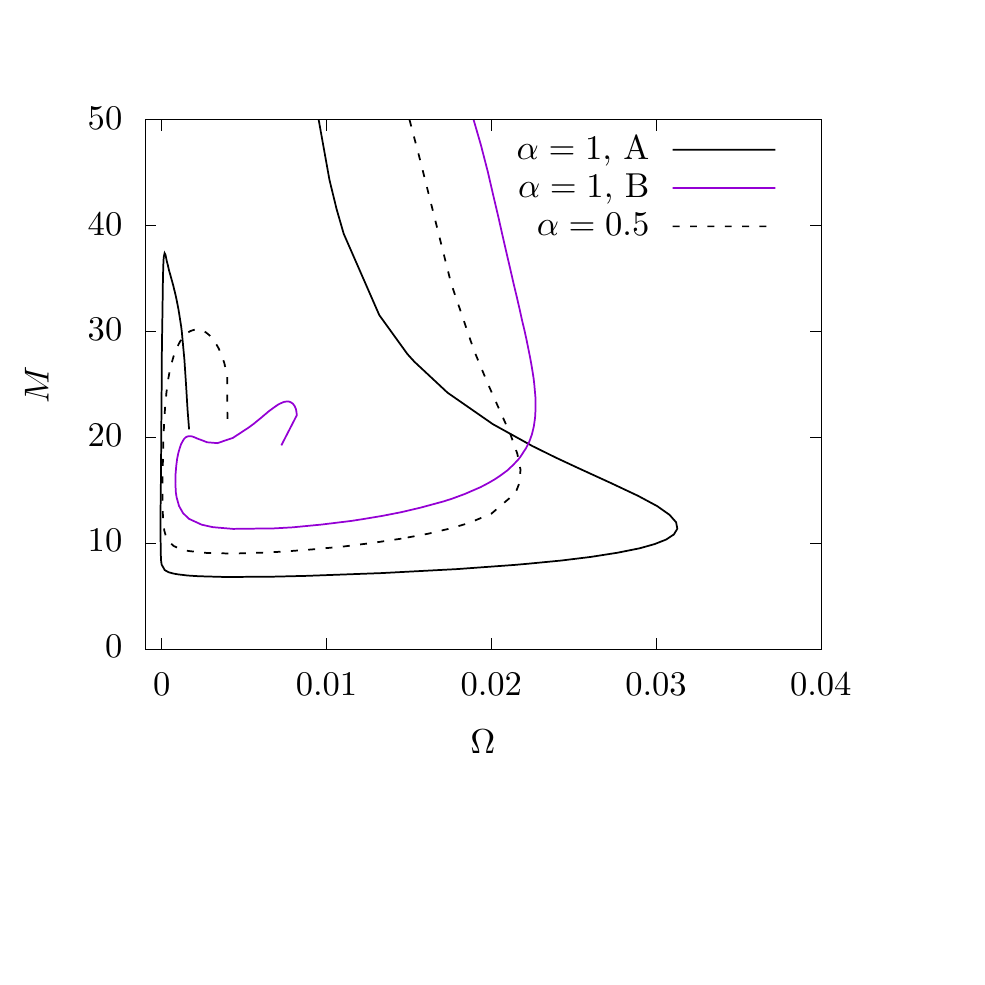}}
{\includegraphics[width=8cm]{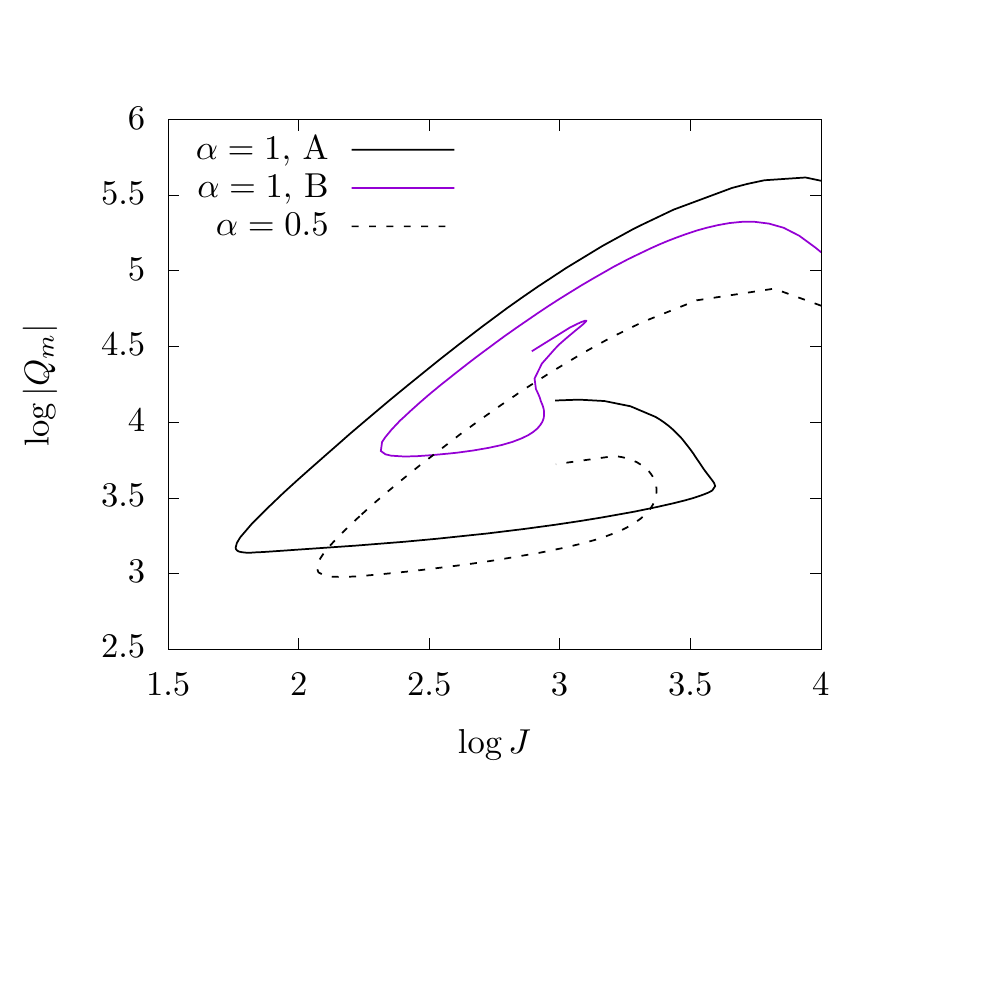}}
\end{center}
\vspace{-2cm}
\caption{{\it Left:} We show the mass $M$ as function of $\Omega$ for $q=0.5$ and $\alpha=1$ (solid) - branch A (black) and
branch B (violet) as well as for $\alpha=0.5$ (dashed). {\it Right:} The magnetic moment $\vert Q_m\vert $ as function of the angular momentum $J$ for the same EMCS boson stars. 
\label{fig:CS_q_05_alpha1_alpha0_5}
}
\end{figure}

When plotting the mass $M$ as function of $\Omega$, see Fig.\ref{fig:CS_q_05_alpha1_alpha0_5} (left)  
we find that the maximal possible value of $\Omega$ increases with increasing $\alpha$ and that on the second branch the solutions exist down to $\Omega \approx 0$ from where a third branch of solutions emerges that shows a sharp increase in mass $M$ on a very small interval of $\Omega$. Reaching a maximal mass, a fourth branch emerges
on which the mass decreases again. We find that the larger $\alpha$ the sharper is the increase of the mass on the third branch.  Comparing the solutions on branch A and branch B for $\alpha=1$ we find that the qualitative dependence
of the mass on $\Omega$ is quite different. In particular, we notice that the qualitative dependence of the solutions
on branch B for $\alpha=1$ seems similar to that of the solutions for $\alpha=0.5$. Finally, we have checked the
dependence of the magnetic moment $\vert Q_m\vert$ on the angular momentum $J$ of the solutions. This is shown in
Fig.\ref{fig:CS_q_05_alpha1_alpha0_5} (right). Interestingly, we find a nearly linear relation between $\log \vert Q_m\vert$ and $\log J$ on the first two branches of solutions, where the first branch has a larger slope than the second.
The third and fourth branch show more complicated behaviour. This relation has been found before
from observations of planets and stars \cite{angular_momentum_magnetic_moment}. While boson stars are 
assumed to be very compact objects and hence rather be comparable in density to neutron stars and white dwarfs, our results suggest that (at least in 5 dimensions) EMCS boson stars have the property that $Q_m$ is proportional to a positive power of $J$, i.e. that $Q_m$ would increase when $J$ increases. This seems to be different for neutron stars and white dwarfs which have $Q_m$ proportional to a negative power of $J$ \cite{angular_momentum_magnetic_moment} and hence the magnetic moment would decrease with increased angular momentum.

\section{Conclusions}
In this paper, we have discussed the construction of charged and rotating boson stars in 5 space-time dimensions. The gauge field dynamics is either of Maxwell type or of Maxwell-Chern-Simons type. These solutions possess electric charge $Q$ equal to $q$ times the Noether charge $N$, where $q$ is the gauge coupling, and sum of the two angular momenta $J$ equal to the Noether charge, i.e. $Q/q=N=J$. The gyromagnetic ratio of the solutions is on the order of unity for boson stars in standard Maxwell gauge field theory, while it can become one order of magnitude larger when the Chern-Simons interaction is added. Moreover, we observe that the presence of the Chern-Simons term leads to the existence of solutions
with radially excited magnetic gauge field component. 
This leads to the reversal of the sign of the magnetic moment and the gyromagnetic ratio, i.e. we find solutions with positive and negative gyromagnetic ratio in the presence of the Chern-Simons term.

For sufficiently compact boson stars we find that the space-time possesses an ergoregion which suggests that these solutions become eventually unstable. The presence of the CS term decreases the radial extention of the ergoregion at fixed $q$ and $\phi'(0)$. 

When considering the relation between the magnetic moment and the angular momentum we find a positive correlation for the solutions on the first and second branch of solutions, i.e. the absolute value of the magnetic moment 
increases with angular momentum. This is different for neutron stars and white dwarfs for which a negative correlation seems to exist, see e.g.
\cite{angular_momentum_magnetic_moment}. Positive correlations are typical for planets and ordinary stars. It would be interesting to investigate this question further in other space-time dimensions.



\begin{thebibliography}{99}



\bibitem{BS1}
D.~J.~Kaup: Klein-Gordon Geon,
Phys. Rev. \textbf{172} (1968), 1331.

\bibitem{BS2}
R.~Friedberg, T.~D.~Lee and Y.~Pang: Mini-soliton stars,
Phys. Rev. D \textbf{35} (1987), 3640. 

\bibitem{BS3}
P.~Jetzer: Boson stars,
Phys. Rept. \textbf{220} (1992), 163.

\bibitem{BS4} 
F.~E.~Schunck and E.~W.~Mielke: General relativistic boson stars,
Class. Quant. Grav. \textbf{20} (2003), R301.


\bibitem{kunz_list_kleihaus} 
B.~Kleihaus, J.~Kunz and M.~List: Rotating boson stars and Q-balls,
Phys. Rev. D \textbf{72} (2005), 064002.

\bibitem{kunz_list_kleihaus_schaffer}
B.~Kleihaus, J.~Kunz, M.~List and I.~Schaffer: Rotating Boson Stars and Q-Balls. II. Negative Parity and Ergoregions,
Phys. Rev. D \textbf{77} (2008), 064025. 

\bibitem{ergoregion_BS} V.~Cardoso, P.~Pani, M.~Cadoni and M.~Cavaglia: Ergoregion instability of ultracompact astrophysical objects,
Phys. Rev. D \textbf{77} (2008), 124044.



\bibitem{Jetzer:1989av}
P.~Jetzer and J.~J.~van der Bij: Charged boson stars,
Phys. Lett. B \textbf{227} (1989), 341.

 \bibitem{Kleihaus:2009kr}
B.~Kleihaus, J.~Kunz, C.~L\"ammerzahl and M.~List: Charged Boson Stars and Black Holes,
Phys. Lett. B \textbf{675} (2009), 102.

\bibitem{Pugliese:2013gsa}
  D.~Pugliese, H.~Quevedo, J.~A.~Rueda H. and R.~Ruffini: On charged boson stars, Phys.\ Rev.\ D {\bf 88} (2013) 024053.




\bibitem{Hartmann:2010pm} B.~Hartmann, B.~Kleihaus, J.~Kunz and M.~List: Rotating Boson Stars in 5 Dimensions,
Phys. Rev. D \textbf{82} (2010), 084022.



\bibitem{Brihaye:2009dx}
  Y.~Brihaye, T.~Caebergs and T.~Delsate: Charged-spinning-gravitating Q-balls,
  arXiv:0907.0913 [gr-qc].


\bibitem{Collodel:2019ohy}
  L.~G.~Collodel, B.~Kleihaus and J.~Kunz: Structure of rotating charged boson stars,
  Phys.\ Rev.\ D {\bf 99} (2019) no.10,  104076. 
	


\bibitem{Kunz:2005ei}
  J.~Kunz and F.~Navarro-Lerida: D=5 Einstein-Maxwell-Chern-Simons black holes,
  Phys.\ Rev.\ Lett.\  {\bf 96} (2006) 081101.
  
\bibitem{kunz:2015kja}
J.~L.~Bl\'azquez-Salcedo, J.~Kunz, F.~Navarro-L\'erida and E.~Radu: Radially excited rotating black holes in Einstein-Maxwell-Chern-Simons theory,
Phys. Rev. D \textbf{92} (2015) no.4, 044025

\bibitem{Kunz:2017pnm}
  J.~Kunz, J.~L.~Blazquez-Salcedo, F.~Navarro-Lerida and E.~Radu:  Einstein-Maxwell-Chern-Simons Black Holes,
  J.\ Phys.\ Conf.\ Ser.\  {\bf 942} (2017) no.1,  012003.

\bibitem{Blazquez:2017kig}
  J.~L.~Blazquez-Salcedo, J.~Kunz, F.~Navarro-Lerida and E.~Radu: New black holes in $D=5$ minimal gauged supergravity: Deformed boundaries and frozen horizons, Phys.\ Rev.\ D {\bf 97} (2018) no.8,  081502.
  
  \bibitem{Brihaye:2018mlv}
  Y.~Brihaye and L.~Ducobu: Spinning-charged-hairy black holes in 5D Einstein gravity,
  Phys.\ Rev.\ D {\bf 98} (2018) no.6,  064034.

\bibitem{mp} 
R.~C.~Myers and M.~J.~Perry: Black Holes in Higher Dimensional Space-Times,
Annals Phys. \textbf{172} (1986), 304.

\bibitem{angular_momentum_magnetic_moment} C.N. Arge, D.J. Mullan, A.Z. Dolginov: Magnetic moments and angular momenta of stars and planets, Astrophy. J. \textbf{433} (1995) 795. 



\bibitem{COLSYS}
 U. Ascher, J. Christiansen, R.~D. Russell: A collocation solver for mixed order systems of boundary value problems,
 Math. of Comp. {\bf 33} (1979) 659;\\
 U. Ascher, J. Christiansen, R.~D. Russell: Collocation software for boundary-value ODEs,
 ACM Trans. {\bf 7} (1981) 209.

\end{thebibliography}
\end{document}